\documentclass[twocolumn,times,tighten]{aastex63}
\newcommand{\uat}[2]{\href{http://astrothesaurus.org/uat/#2}{#1 (#2)}}

\received{2020 November 17}
\revised{2020 December 22}
\accepted{2021 January 11}
\acceptjournal{The Astrophysical Journal Letters}
\reportnum{\url{https://arxiv.org/abs/2011.08653}}

\shorttitle{\scriptsize{\textsc{The Astrophysical Journal Letters, 00:000000 (10pp), 2021 Month Day}}}
\shortauthors{Coti Zelati et al.}

\begin{document}

\title{The New Magnetar \src\ in Outburst}


 
\author[0000-0001-7611-1581]{F. Coti Zelati}
\affiliation{Institute of Space Sciences (ICE, CSIC), Campus UAB, Carrer de Can Magrans s/n, E-08193, Barcelona, Spain; \href{mailto:cotizelati@ice.csic.es}{cotizelati@ice.csic.es}}
\affiliation{Institut d'Estudis Espacials de Catalunya (IEEC), Carrer Gran Capit\`a 2--4, E-08034 Barcelona, Spain} 

\author[0000-0001-8785-5922]{A. Borghese}
\affiliation{Institute of Space Sciences (ICE, CSIC), Campus UAB, Carrer de Can Magrans s/n, E-08193, Barcelona, Spain; \href{mailto:cotizelati@ice.csic.es}{cotizelati@ice.csic.es}}
\affiliation{Institut d'Estudis Espacials de Catalunya (IEEC), Carrer Gran Capit\`a 2--4, E-08034 Barcelona, Spain} 

\author[0000-0001-5480-6438]{G. L. Israel}
\affiliation{INAF--Osservatorio Astronomico di Roma, via Frascati 33, I-00078 Monteporzio Catone, Italy} 

\author[0000-0003-2177-6388]{N. Rea}
\affiliation{Institute of Space Sciences (ICE, CSIC), Campus UAB, Carrer de Can Magrans s/n, E-08193, Barcelona, Spain; \href{mailto:cotizelati@ice.csic.es}{cotizelati@ice.csic.es}}
\affiliation{Institut d'Estudis Espacials de Catalunya (IEEC), Carrer Gran Capit\`a 2--4, E-08034 Barcelona, Spain} 

\author[0000-0003-4849-5092]{P. Esposito}
\affiliation{Scuola Universitaria Superiore IUSS Pavia, Palazzo del Broletto, piazza della Vittoria 15, I-27100 Pavia, Italy}
\affiliation{INAF--Istituto di Astrofisica Spaziale e Fisica Cosmica di Milano, via A.\,Corti 12, I-20133 Milano, Italy}

\author[0000-0001-7397-8091]{M. Pilia}
\affiliation{INAF--Osservatorio Astronomico di Cagliari, Via della Scienza 5, I-09047 Selargius, Italy}

\author[0000-0002-8265-4344]{M. Burgay}
\affiliation{INAF--Osservatorio Astronomico di Cagliari, Via della Scienza 5, I-09047 Selargius, Italy}

\author[0000-0001-5902-3731]{A. Possenti}
\affiliation{INAF--Osservatorio Astronomico di Cagliari, Via della Scienza 5, I-09047 Selargius, Italy}
\affiliation{Department of Physics, Universit\`a di Cagliari, S.P. Monserrato-Sestu km 0,700, I-09042 Monserrato, Italy}

\author[0000-0002-5924-3141]{A. Corongiu}
\affiliation{INAF--Osservatorio Astronomico di Cagliari, Via della Scienza 5, I-09047 Selargius, Italy}

\author[0000-0001-6762-2638]{A. Ridolfi}
\affiliation{INAF--Osservatorio Astronomico di Cagliari, Via della Scienza 5, I-09047 Selargius, Italy}
\affiliation{Max Planck Institute f\"ur Radioastronomie, Auf dem H\"ugel 69, D-53121 Bonn, Germany}

\author[0000-0003-0554-7286]{C. Dehman}
\affiliation{Institute of Space Sciences (ICE, CSIC), Campus UAB, Carrer de Can Magrans s/n, E-08193, Barcelona, Spain; \href{mailto:cotizelati@ice.csic.es}{cotizelati@ice.csic.es}}
\affiliation{Institut d'Estudis Espacials de Catalunya (IEEC), Carrer Gran Capit\`a 2--4, E-08034 Barcelona, Spain} 

\author[0000-0001-7795-6850]{D. Vigan\`o}
\affiliation{Institute of Space Sciences (ICE, CSIC), Campus UAB, Carrer de Can Magrans s/n, E-08193, Barcelona, Spain; \href{mailto:cotizelati@ice.csic.es}{cotizelati@ice.csic.es}}
\affiliation{Institut d'Estudis Espacials de Catalunya (IEEC), Carrer Gran Capit\`a 2--4, E-08034 Barcelona, Spain} 

\author[0000-0003-3977-8760]{R. Turolla}
\affiliation{Dipartimento di Fisica e Astronomia ``Galileo Galilei'', Universit\`a di Padova, via F. Marzolo 8, I-35131 Padova, Italy}
\affiliation{Mullard Space Science Laboratory, University College London, Holmbury St. Mary, Dorking, Surrey RH5 6NT, UK}

\author[0000-0001-5326-880X]{S. Zane}
\affiliation{Mullard Space Science Laboratory, University College London, Holmbury St. Mary, Dorking, Surrey RH5 6NT, UK}

\author[0000-0002-6038-1090]{A. Tiengo}
\affiliation{Scuola Universitaria Superiore IUSS Pavia, Palazzo del Broletto, piazza della Vittoria 15, I-27100 Pavia, Italy}
\affiliation{INAF--Istituto di Astrofisica Spaziale e Fisica Cosmica di Milano, via A.\,Corti 12, I-20133 Milano, Italy}
\affiliation{Istituto Nazionale di Fisica Nucleare (INFN), Sezione di Pavia, via A.\,Bassi 6, I-27100 Pavia, Italy}

\author[0000-0002-4553-655X]{E. F. Keane} 
\affiliation{SKA Organisation, Jodrell Bank, Macclesfield, Cheshire, SK11 9FT, UK}
\affiliation{Centre for Astronomy, School of Physics, National University of Ireland Galway, University Road, Galway, H91 TK33, Ireland}

\def\xmm {\emph{XMM-Newton}}
\def\cxo {\emph{Chandra}}
\def\nustar {\emph{NuSTAR}}
\def\rst {\emph{ROSAT}}
\def\swift {\emph{Swift}}
\def\nicer {\emph{NICER}}
\def\pks {Parkes}

\def\flux {\mbox{erg\,cm$^{-2}$\,s$^{-1}$}}
\def\lum {\mbox{erg\,s$^{-1}$}}
\def\nh {N_{\rm H}}
\def\kms  {\rm \ km \, s^{-1}}
\def\cms  {\rm \ cm \, s^{-1}}
\def\gs   {\rm \ g  \, s^{-1}}
\def\cmtre {\rm \ cm^{-3}}
\def\cmdue {\rm \ cm$^{-2}$}
\def\ss {\mbox{s\,s$^{-1}$}}
\def\chisq {$\chi ^{2}$}
\def\rchisq {$\chi_{r} ^{2}$}

\def\arc{\mbox{$^{\prime\prime}$}}
\def\arcmin{\mbox{$^{\prime}$}}
\def\deg{\mbox{$^{\circ}$}}

\def\rsun {~R_{\odot}}
\def\msun {~M_{\odot}}
\def\mdotav {\langle \dot {M}\rangle }

\def\uu {4U\,0142$+$614}
\def\ee {1E\,1048.1$-$5937}
\def\kes {1E\,1841$-$045}
\def\aa {1E\,1547$-$5408}
\def\axj {AX\,J1844$-$0258}
\def\rxs {1RXS\,J1708$-$4009}
\def\xte{XTE\,J1810$-$197}
\def\smc{CXOU\,J0100$-$7211}
\def\wes{CXOU\,J1647$-$4552}
\def\ea {1E\,2259$+$586}
\def\ctb{CXOU\,J171405.7$-$381031}
\def\sgra{SGR\,1806$-$20}
\def\sgrb{SGR\,1900$+$14}
\def\sgrd{SGR\,1627$-$41}
\def\sgre{SGR\,0501$+$4516}
\def\sgrf{SGR\,1935+2154}
\def\lowba{SGR\,0418$+$5729}
\def\sgrg{SGR\,1833$-$0832}
\def\lowbb{Swift\,J1822.3$-$1606}
\def\galmag{PSR\,J1745$-$2900}
\def\sgras{Sgr\,A$^{\star}$}
\def\sgrh{SGR\,1801$-$21}
\def\sgri{SGR\,2013$+$34}
\def\psr{PSR\,1622$-$4950}
\def\hbpsr{PSR\,J1846$-$0258}
\def\radiohb{PSR\,J1119$-$6127}
\def\coronamag{Swift\,J1818.0$-$1607}

\def\src {\mbox{SGR\,J1830$-$0645}}

\begin{abstract}
The detection of a short hard X-ray burst and an associated bright soft X-ray source by the \swift\ satellite in 2020 October heralded a new magnetar in outburst, \src. Pulsations at a period of $\sim$10.4\,s were detected in prompt follow-up X-ray observations. We present here the analysis of the \swift/BAT burst, of \xmm\ and the {\em Nuclear Spectroscopic Telescope Array} observations performed at the outburst peak, and of a \swift/XRT monitoring campaign over the subsequent month. The burst was single-peaked, lasted $\sim$6\,ms, and released a fluence of $\approx$ $5\times10^{-9}$\,erg\,cm$^{-2}$ (15--50\,keV). The  spectrum of the X-ray source at the outburst peak was well described by an absorbed double-blackbody model plus a power-law component detectable up to $\sim$25\,keV. The unabsorbed X-ray flux decreased from $\sim5\times10^{-11}$ to $\sim2.5\times10^{-11}$\,\flux\ one month later (0.3--10\,keV). Based on our timing analysis, we estimate a dipolar magnetic field $\approx5.5\times10^{14}$\,G at pole, a spin-down luminosity $\approx2.4\times10^{32}$\,\lum, and a characteristic age $\approx$24\,kyr. The spin modulation pattern appears highly pulsed in the soft X-ray band, and becomes smoother at higher energies. Several short X-ray bursts were detected during our campaign. No evidence for periodic or single-pulse emission was found at radio frequencies in observations performed with the Sardinia Radio Telescope and Parkes. According to magneto-thermal evolutionary models, the real age of \src\ is close to the characteristic age, and the dipolar magnetic field at birth was slightly larger, $\sim$10$^{15}$\,G.
\end{abstract}

\section{Introduction} \label{sec:intro}
The term ``magnetar'' was coined almost three decades ago to identify isolated neutron stars (NSs) ultimately powered by the dissipation of their own magnetic energy, which usually implies that they are endowed with huge magnetic fields, up to $\sim$10$^{15}$\,G \citep{duncan92}. A large fraction of the $\sim$30 magnetars known to date \citep{olausen14} have been discovered just over the past two decades, through their distinctive high-energy phenomenology: bursts of X-ray/gamma-ray emission and/or enhancements of their persistent X-ray luminosity, dubbed ``outbursts'' \citep[see][]{kaspi17,esposito21}. The bursts are comparatively brief episodes lasting from milliseconds to hundreds of seconds, and reaching X-ray peak luminosities within the interval $10^{39}-10^{47}$\,\lum\ \citep[e.g.,][]{collazzi15}. The outbursts are instead long-lasting events where the X-ray luminosity firstly rises to values in the range of $10^{34}-10^{36}$\,\lum, and then decreases on timescales that can be as long as years \citep{cotizelati18}. Remarkably, in a couple of magnetars the post-outburst luminosity was found to differ from the pre-outburst long-term persistent level \citep{ybk17,cotizelati20}.

On 2020 October 10 at 14:49:24 UT, the Burst Alert Telescope (BAT) on board the {\em Neil Gehrels Swift Observatory} triggered on a short, hard X-ray burst \citep{page20}. Prompt observations with the \swift\ X-ray Telescope (XRT) localized a new uncatalogued X-ray source, \src.\footnote{\citet{tohuvavohu20} reported the discovery in an offline search of the BAT data of another burst from \src\ which, however, did not result in a detector trigger.} An X-ray periodic signal at $\sim$10.4\,s was detected in the XRT data \citep{gogus20a} and confirmed later by observations with the Neutron Star Interior Composition Explorer (\nicer) \citep{younes20}. The burst properties, the periodicity detected in the prompt follow-up observations, and the proximity of the source to the Galactic plane (Galactic latitude $b\sim1.5$\deg) point to a newly discovered magnetar in outburst.  

This Letter reports on (i) the properties of the X-ray bursts detected from \src\ by the \swift/BAT; (ii) quasi-simultaneous observations with \xmm\ and the Nuclear Spectroscopic Telescope Array (\nustar) performed within $\sim$2 days after the first BAT burst; (iii) a \swift/XRT monitoring campaign covering the first month since the outburst onset; (iv) a search for short bursts in the X-ray time series (Section\,\ref{sec:X}); and (v) radio observations with the Sardinia Radio Telescope and \pks\ (Section\,\ref{sec:radio}). Discussion and conclusions follow (Section\,\ref{sec:discussion}).

\begin{deluxetable*}{ccccccc}[htb]
\tablecaption{Observation Log, X-Ray Fluxes, and Limits on Pulsed Radio Emission
\label{tab:observations}}
\tabletypesize{\scriptsize}
\tablecolumns{7}
\tablenum{1}
\tablewidth{0pt}
\tablehead{
\colhead{X-Ray Instrument\tablenotemark{{\scriptsize a}}} &
\colhead{Obs.ID} &
\colhead{Start} &
\colhead{Stop} & 
\colhead{Exposure} & 
\colhead{Net Count Rate\tablenotemark{{\scriptsize b}}} &
\colhead{Flux (Obs / Unabs)\tablenotemark{{\scriptsize c}} } \\  [-0.3cm]
\colhead{} &
\colhead{} & 
\multicolumn2c{YYYY Mmm DD hh:mm:ss (TT)} &
\colhead{(ks)} & 
\colhead{(counts\,s$^{-1}$)} &
\colhead{($\times 10^{-11}$\,\flux)}
}
\startdata
\rst/PSPC & rp500012n00 & 1991 Apr 3 20:34:56 & 1991 Apr 3 23:04:12 & 4.5 & $<$0.008 & $<$0.008 / $<$0.15 \\
\hline
\swift/XRT (PC) & 00999571000 & 2020 Oct 10 14:51:07  & 2020 Oct 10 16:30:29 & 1.7 & 0.47$\pm$0.02\tablenotemark{{\scriptsize d}} & (4.0$\pm$0.2) / (5.0$\pm$0.2)  \\
\swift/XRT (PC) & 00999571001 & 2020 Oct 10 17:53:14  & 2020 Oct 10 22:43:35 & 4.5 & 0.46$\pm$0.01 & (3.8$\pm$0.1) / (4.6$\pm$0.1)  \\
\xmm/EPIC-pn (SW) & 0872390501 & 2020 Oct 11 21:07:13 & 2020 Oct 12 03:41:12 & 16.5\tablenotemark{{\scriptsize e}} & 7.09$\pm$0.02 & (3.96$\pm$0.02) / (5.11$\pm$0.02)  \\
\nustar/FPMA & 90601331002 & 2020 Oct 12 07:46:09   & 2020 Oct 12 23:26:09 & 29.6 & 0.709$\pm$0.005 & (3.96$\pm$0.02) / (5.11$\pm$0.02)  \\
\nustar/FPMB & 90601331002 & 2020 Oct 12 07:46:09   & 2020 Oct 12 23:26:09 & 29.4 & 0.666$\pm$0.005 & (3.96$\pm$0.02) / (5.11$\pm$0.02)  \\
\swift/XRT (WT) & 00999571002 & 2020 Oct 15 01:15:03   & 2020 Oct 15 20:35:55  & 2.3 & 0.69$\pm$0.02  & (4.1$\pm$0.2) / (5.1$\pm$0.2)  \\
\swift/XRT (WT) & 00999571003 & 2020 Oct 16 10:35:46   & 2020 Oct 17 19:04:56  & 2.6 & 0.69$\pm$0.02  & (3.7$\pm$0.2) / (4.6$\pm$0.2)  \\
\swift/XRT (WT) & 00999571004 & 2020 Oct 19 04:18:46   & 2020 Oct 19 22:03:56  & 3.3 & 0.66$\pm$0.02 & (3.5$\pm$0.1) / (4.4$\pm$0.1)  \\
\swift/XRT (WT) & 00999571005 & 2020 Oct 23 11:33:32   & 2020 Oct 23 14:51:44  & 0.9 & 0.55$\pm$0.03 & (3.3$\pm$0.5) / (4.2$\pm$0.5)  \\
\swift/XRT (WT) & 00013840001 & 2020 Oct 29 04:35:55   & 2020 Oct 29 22:34:56 & 2.9 & 0.51$\pm$0.02 & (2.9$\pm$0.1) / (3.6$\pm$0.2)  \\
\swift/XRT (WT) & 00013840002 & 2020 Nov 1 15:23:43   & 2020 Nov 2 11:00:56 & 4.1 & 0.42$\pm$0.01 & (3.0$\pm$0.2) / (3.8$\pm$0.2)  \\
\swift/XRT (WT) & 00013840003 & 2020 Nov 4 10:24:01   & 2020 Nov 4 23:19:50 & 2.8 & 0.47$\pm$0.01 & (2.5$\pm$0.1) / (3.2$\pm$0.1)  \\
\swift/XRT (PC) & 01004219000 & 2020 Nov 5 02:24:07   & 2020 Nov 5 02:52:52 & 1.7 & 0.34$\pm$0.01 & (2.6$\pm$0.3) / (3.1$\pm$0.3)  \\
\swift/XRT (WT) & 00013840004 & 2020 Nov 6 11:42:27   & 2020 Nov 6 18:19:56  & 3.8 & 0.48$\pm$0.01 & (2.6$\pm$0.1) / (3.3$\pm$0.1)  \\
\swift/XRT (WT) & 00013840005 & 2020 Nov 10 05:01:02   & 2020 Nov 10 10:07:56 & 3.9 & 0.44$\pm$0.01 & (2.3$\pm$0.1) / (2.9$\pm$0.1)  \\
\swift/XRT (PC) & 01005428000 & 2020 Nov 11 09:46:28   & 2020 Nov 11 10:15:10 & 1.7 & 0.37$\pm$0.01 & (3.1$\pm$0.5) / (3.6$\pm$0.5)  \\
\swift/XRT (WT) & 00013840006 & 2020 Nov 13 01:34:21  & 2020 Nov 13 13:09:56 & 3.6 & 0.33$\pm$0.01 & (2.0$\pm$0.1) / (2.6$\pm$0.1)  \\  
\hline 
\hline
\colhead{Radio Instrument}  & 
\colhead{Frequency} &
\colhead{Bandwidth} &
\colhead{Start}  &
\colhead{Exposure} &
\colhead{Pulsed Flux Density\tablenotemark{{\scriptsize f}}} \\ [-0.3cm]
\colhead{} & 
\colhead{(GHz)} &
\colhead{(MHz)} &
\colhead{YYYY Mmm DD hh:mm:ss (TT)}  &
\colhead{(hr)} &
\colhead{($\mu$Jy)} \\ 
\hline 
SRT 		&  1.5 & 500  & 2020 Oct 11 16:20:19   & 2.7 & $<$90 	\\ 
Parkes 	    &  0.96 & 512 & 2020 Oct 12 08:05:33   & 2.9 & $<$77   \\
Parkes 	    &  1.6 & 768 & 2020 Oct 12 08:05:33   & 2.9 & $<$39   \\
Parkes 	    &  2.4 & 768 & 2020 Oct 12 08:05:33   & 2.9 & $<$34   \\
Parkes 	    &  3.4 & 1280 & 2020 Oct 12 08:05:33   & 2.9 & $<$25   \\
SRT 		&  6.8 & 900  & 2020 Oct 21 11:59:29   & 2.0 & $<$45   \\
SRT 		&  6.8 & 900  & 2020 Oct 21 14:05:39   & 0.7 & $<$77   \\
SRT 		&  6.8 & 900  & 2020 Oct 21 15:05:09   & 4.9 & $<$29   \\
SRT 		&  6.8 & 900  & 2020 Oct 30 11:02:30   & 7.8 & $<$23   \\
SRT 		&  6.8 & 900  & 2020 Nov 6 10:30:00   & 7.1 & $<$24   \\
\enddata
\tablecomments{
\vspace{-0.25cm}
\tablenotetext{a}{The instrumental setup is indicated in brackets: PC = photon counting, SW = small window, WT = windowed timing.} \vspace{-0.25cm}
\tablenotetext{b}{The count rate is in the 0.3--10\,keV energy range, except for \rst\ (0.1--2.4\,keV) and \nustar\ (3--25\,keV). The upper limit is quoted at 3$\sigma$ c.l.} \vspace{-0.25cm}
\tablenotetext{c}{The flux is in the 0.3--10\,keV energy range. The upper limits are quoted at 3$\sigma$ c.l., and are computed assuming an absorbed blackbody spectrum with $\nh = 1.07\times10^{22}$\,cm$^{-2}$ and $kT_{\rm BB}=0.15$\,keV.} \vspace{-0.25cm}
\tablenotetext{d}{Corrected for pile-up.} \vspace{-0.25cm}
\tablenotetext{e}{Corrected for dead-time.} \vspace{-0.25cm}
\tablenotetext{f}{Upper limits are computed using the radiometer equation \citep{lorimer04}, assuming a pulse duty cycle of 10\%.}  \vspace{-0.8cm}
}
\end{deluxetable*}

\section{X-Ray Emission}\label{sec:X}
\subsection{Observations and Data Analysis} 
Table\,\ref{tab:observations} reports a log of the X-ray observations. Data reduction was performed using tools incorporated in {\sc heasoft} (v.6.28) and the Science Analysis Software (v.19) with the latest calibration files. Photon arrival times were barycentered using the \cxo\ position \citep{gogus20b}, R.A. = 18$^\mathrm{h}$30$^\mathrm{m}$41$\fs$64, decl. = --06$^{\circ}$45$^{\prime}$16$\farcs$9 (J2000.0; uncertainty of 0$\farcs$8 at 90\% c.l.\footnote{The uncertainty is dominated by the satellite absolute positional accuracy. See \url{https://cxc.harvard.edu/cal/ASPECT/celmon/}.}) and the JPL planetary ephemeris DE-405. The spectral analysis was performed using the {\sc xspec} package \citep{arnaud96}, adopting the {\sc Tbabs} model \citep{wilms00} to describe the interstellar absorption. Hereafter, all uncertainties are quoted at 1$\sigma$ c.l.

\subsubsection{Swift}\label{sec:swift}
For the burst that revealed \src\ by alerting the BAT and two other events detected on 2020 November 5 and 11, we extracted mask-tagged light curves and spectra using standard tools in the {\sc ftools} software package. \src\ was observed 14 times with the XRT \citep{burrows05} configured either in photon counting (PC; timing resolution of 2.5\,s), or windowed timing (WT; 1.8\,ms) mode. The source photons were extracted from a circular region of radius 20 pixels (1 pixel=2$\farcs$36). Background events were collected from a circle of the same size for WT-mode data and an annulus with radii 40--80 pixels for PC-mode data. In the first pointing (obs.ID 00999571000), photons within the inner 4 pixels of the source point spread function were removed to minimize pile-up effects.

\begin{figure*}
\centering
\resizebox{\hsize}{!}{\includegraphics[angle=0]{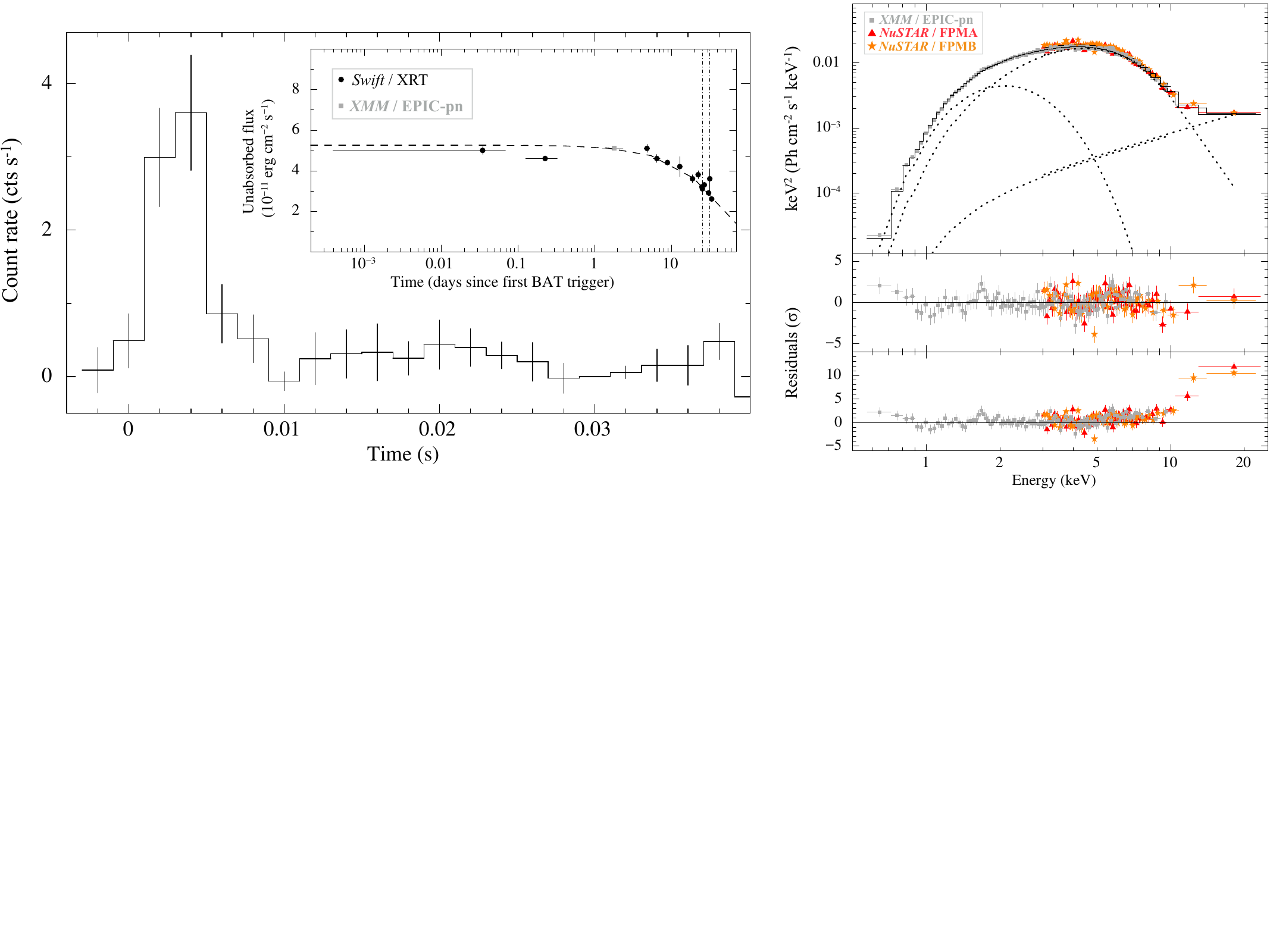}}
\vspace{-7.2cm}
\caption{\label{fig:spectroscopy} {\it Left}: \swift/BAT light curve of the burst that led to the discovery of \src\ (15--50\,keV energy range; time bin: 2\,ms; the start time is arbitrary). The inset shows the evolution of the 0.3--10\,keV unabsorbed flux of \src\ measured over a baseline of about one month since the burst trigger (MJD 59,132.6176). The dashed line marks the best-fitting exponential function. The vertical dotted--dashed lines mark the epochs of the other two bursts detected by the BAT (MJD 59,158.0986 and 59,164.4071; \citealt{gropp20a,gropp20b}). {\it Right, Top}: unfolded spectrum extracted from the \xmm/EPIC-pn (gray), \nustar/FPMA (red) and \nustar/FPMB (orange) data. The solid line indicates the best-fitting model, the dotted lines show the contribution of the different spectral components. {\it Middle}: post-fit residuals. {\it Bottom}: residuals obtained after removing the power-law component from the model.}
\end{figure*}

\subsubsection{XMM--Newton}\label{sec:xmm}
\src\ was observed with the European Photon Imaging Cameras (EPIC) on board \xmm\ on 2020 October 11--12, for an exposure time of 23.6\,ks. The EPIC-pn \citep{struder01} and the MOS \citep{turner01} cameras were operating in Small Window mode (SW; timing resolutions of 5.7\,ms and 0.3\,s, respectively). Here, we consider only the data acquired with the EPIC-pn, which provides the data set with the highest counting statistics owing to its larger effective area compared to the MOS cameras.

Raw data were processed following standard analysis procedures. No periods of high background activity were detected. 
The source events were selected from a circle with a radius of 40\arc\ and the background counts were accumulated from a closeby circle of the same size. The response matrices and ancillary files were generated through the {\sc rmfgen} and {\sc arfgen} tools, respectively. Background-subtracted and exposure-corrected light curves were extracted using the {\sc epiclccorr} task.

\subsubsection{NuSTAR}\label{sec:nustar}
\nustar\ \citep{harrison13} observed \src\ on 2020 October 12 for an effective exposure time of 29.6\,ks. We created cleaned event files and filtered out passages through the South Atlantic Anomaly using the tool {\sc nupipeline} with default options. For both focal plane modules (FPMA and FPMB), we collected the source counts within a circle of radius 100\arc\ and estimated the background from four circles of the same size located on all detectors using the {\sc nuskybgd} pipeline \citep{wik14}. The source was detected up to an energy of $\sim$25\,keV in both FPMs. We ran the tool {\sc nuproducts} to extract light curves and spectra, and to generate response files.

\begin{figure*}
\centering
\resizebox{\hsize}{!}{\includegraphics[angle=0]{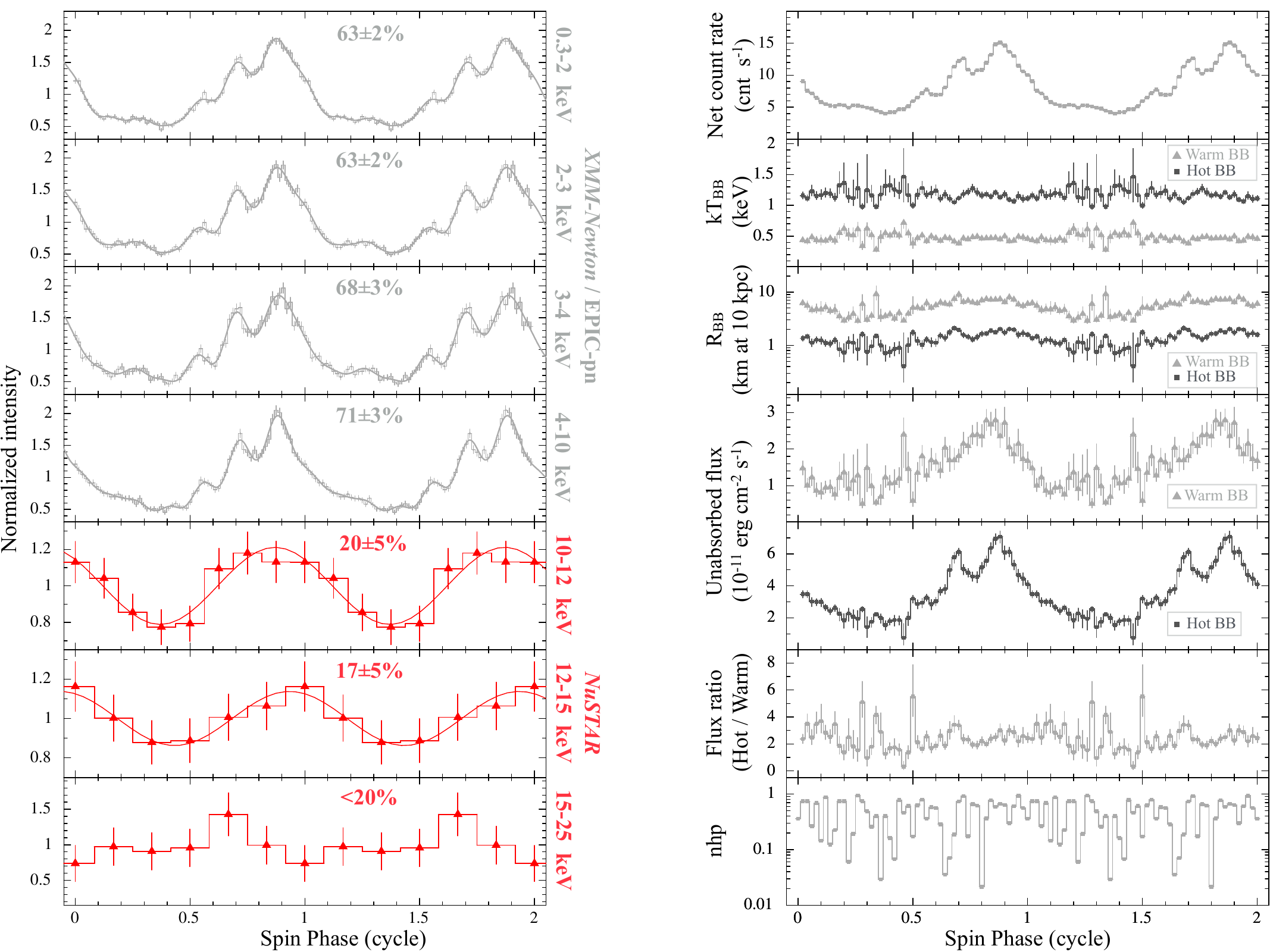}}
\vspace{-0.5cm}
\caption{\label{fig:timing} {\it Left}: energy-resolved background-subtracted pulse profiles of \src\ extracted from \xmm/EPIC-pn (gray) and \nustar\ (red) data. The best-fitting models obtained by using seven sinusoidal components (fundamental plus harmonics) for EPIC-pn and a single sinusoidal component (fundamental) for \nustar\ are indicated with solid lines. The corresponding pulsed fraction values (or the 3$\sigma$ upper limit for the 15--25\,keV range) are reported in each panel. {\it Right}: results of the spin phase-resolved spectroscopy of EPIC-pn data in the 0.3--10\,keV range. From top to bottom: background-subtracted pulse profile; blackbody temperature, radius (assuming a distance of 10\,kpc) and 0.3--10\,keV unabsorbed flux for the warm (light gray) and hot (dark gray) components; hot-to-warm blackbody unabsorbed flux ratio; nhp values derived from the \chisq\ and the d.o.f. of the fit of each spectrum. All uncertainties are at 1$\sigma$ c.l.}
\end{figure*}

\subsection{Results}\label{results}
\subsubsection{BAT Bursts Properties}\label{sec:BATbursts}
Figure\,\ref{fig:spectroscopy} shows the time evolution of the burst detected on 2020 October 10 in the 15--50\,keV band (no emission is detected at higher energies). The event was single-peaked and had a total duration and a $T_{90}$ duration\footnote{Time interval containing 90\% of the counts.}, as computed from a Bayesian blocks analysis with the {\sc battblocks} tool, of $6\pm1$\,ms and $5\pm1$\,ms, respectively. The spectrum extracted from the whole interval, similarly to standard magnetar bursts seen at hard X-rays, can be described in the 15--50\,keV band by simple models, such as a blackbody with temperature $kT=8.0^{+1.6}_{-1.2}$\,keV (with reduced \chisq\ of $\chi^2_r=1.05$ for 14 degrees of freedom (d.o.f.)) or a power law with photon index $\Gamma=1.9\pm0.5$ ($\chi^2_r=1.12$ for 14 d.o.f.). For the blackbody model, the average flux was $7.6^{+0.4}_{-3.3}\times10^{-7}$\,\flux, corresponding to a fluence of $\sim$ $4.6\times10^{-9}$\,erg\,cm$^{-2}$ (15--50\,keV).
Two other bursts resulted in BAT triggers during our campaign \citep[see][]{palmer20}.
The second burst detected on 2020 November 5 was longer ($25\pm5$\,ms, $T_{90}=21\pm5$\,ms) and harder, being visible in the light curve up to $\approx$150\,keV. Also in this case, the adoption of a blackbody model resulted in the best fit, with $\chi^2_r=0.83$ for 56 d.o.f., and $kT=9.8\pm0.7$\,keV. The average flux was $(1.3\pm0.1)\times10^{-6}$\,\flux, while the fluence was $\sim$$3.2\times10^{-8}$\,erg\,cm$^{-2}$ (15--150\,keV). The burst that triggered BAT on 2020 November 11 was intermediate in hardness between the other two events, and lasted $26\pm6$\,ms ($T_{90}=21\pm7$\,ms). For the blackbody model ($kT=9.1\pm0.6$\,keV; $\chi^2_r=0.67$ for 56 d.o.f.), the average flux was $(8.1\pm0.7)\times10^{-7}$\,\flux\ and the fluence $\sim$$2.2\times10^{-8}$\,erg\,cm$^{-2}$ (15--150\,keV).

\subsubsection{X-Ray Monitoring and Archival Observations}\label{sec:xrayanalysis}
Firstly, we fit the broadband spectrum extracted from the quasi-simultaneous EPIC-pn and \nustar\ data sets using models comprising different combinations of blackbody and power-law components. We included a constant term in the fits to account for inter-calibration uncertainties between the three instruments, deriving a mismatch of $<$5\% for all the tested models. The best description of the data was provided by an absorbed double-blackbody model plus a power-law component accounting for emission at energies above $\sim$12\,keV (Figure\,\ref{fig:spectroscopy}), giving \rchisq\ = 1.10 for 376 d.o.f. and a null hypothesis probability nhp$\simeq$0.10. Best-fitting parameters were absorption column density $\nh\ = (1.07 \pm 0.02)\times10^{22}$\,\cmdue, blackbody temperatures and radii $kT_{\rm BB,W} =  0.45\pm0.01$\,keV, $R_{\rm BB,W} =  5.6\pm0.3$\,km for the warm component and $kT_{\rm BB,H} = 1.11\pm0.01$\,keV, $R_{\rm BB,H} = 1.53\pm0.03$\,km for the hot component, respectively (assuming a distance of 10\,kpc; see Section\,\ref{sec:discussion}), and $\Gamma=0.9\pm0.3$. The observed and unabsorbed fluxes were $(4.33\pm0.03)\times10^{-11}$ and $(5.54\pm0.03)\times10^{-11}$\,\flux\ in the 0.3--25\,keV energy range, with a fractional contribution of the power-law component of $\simeq$6\% in the same band.

Then, we fit an absorbed double-blackbody model to all \swift/XRT data, fixing the column density at $\nh\ = 1.07\times10^{22}$\,\cmdue, and allowing all other parameters to vary across the data sets. The observed and unabsorbed fluxes derived from the above model are reported in Table\,\ref{tab:observations}, while the evolution of the unabsorbed flux is shown in the inset of the left panel in Figure\,\ref{fig:spectroscopy}. The unabsorbed flux decreased by a factor of $\sim$2 along our campaign, from $\sim5\times10^{-11}$\,\flux\ at peak to $\sim2.5\times10^{-11}$\,\flux\ about one month later (0.3--10\,keV). Its time evolution can be adequately described so far by an exponential function with $e$-folding time $\tau=55\pm2$\,days (\rchisq = 1.42 for 13 d.o.f.; Figure\,\ref{fig:spectroscopy}).

\src\ was in the field of view of \rst/PSPC in a pointing performed on 1991 April 3 (see Table~\ref{tab:observations}). The source is not detected, and we set an upper limit on the net count rate of 0.008 counts\,s$^{-1}$ (3$\sigma$ c.l.; 0.1--2.4\,keV). Assuming emission from the entire surface ($R_{{\rm NS}}=10$--$15$\,km) and a source distance of 10\,kpc, we estimate that the blackbody temperature should be $\lesssim$0.15\,keV to be consistent with the above limit. The corresponding limits on the observed and unabsorbed flux are $F_{{\rm X,obs}}<8\times10^{-14}$ \flux\ and  $F_{{\rm X,unabs}}<1.5\times10^{-12}$ \flux\ (0.3--10\,keV).

\subsubsection{Timing Analysis and Phase-resolved Spectroscopy}\label{sec:timing}
The  EPIC-pn, \swift/XRT and \nustar\ source event files were used to study the magnetar timing properties. We built up a phase-coherent timing solution starting from the period $P = 10.41572(1)$\,s inferred from the EPIC-pn data sets (those with the largest statistics), and employing a phase-fitting technique. Within a baseline of about 34 days (2020 October 10 -- November 13), we clearly detected a first period derivative component in the signal phase history, and derived the following timing solution: $P = 10.415724(1)$\,s, $\dot{P} = 7(1)\times10^{-12}$\,s\,s$^{-1}$ at the reference epoch $T_0 = 59,133.0$\,MJD. The $\dot{P}$ value is in agreement with that derived by \citet{ray20} using \nicer\ data sets. Our timing solution implies an rms variability of $\sim$145\,ms, corresponding to a timing noise level $<$2\%, similar to the range of values observed in other isolated NSs. We set a 3$\sigma$ upper limit on the second period derivative of $|\ddot{P}| < 2 \times10^{-17}$\,s\,s$^{-2}$. 

Figure\,\ref{fig:timing} shows the background-subtracted light curves extracted from the EPIC-pn and \nustar\ data sets over different energy intervals, folded using the above ephemeris. Pulsed emission was detected up to an energy of $\sim$15\,keV. The pulse profile displays an apparent complex morphology below 10\,keV, with a pronounced dip close to the main peak, a weaker peak in the rising part of the profile, and small-amplitude structures at minimum (seemingly less prominent above $\sim$4\,keV). 
The profile appears to evolve to a relatively simpler shape at higher energies in the 12--15\,keV energy range. 
The pulse peak in this range lags the main peak observed at lower energies by 0.11$\pm$0.06 spin phase cycles. The background-subtracted peak-to-peak semiamplitude increases from (63$\pm$2)\% below 3\,keV to (71$\pm$3)\% in the 4--10\,keV band, and drops to $\sim$20\% in the 10--12 and 12--15\,keV ranges. We set a 3$\sigma$ upper limit of 20\% in the 15--25\,keV range (Figure\,\ref{fig:timing}). The marked changes in the pulse profile amplitude at energies where the power-law spectral component dominates the source emission (Figure\,\ref{fig:spectroscopy}) indicate that the power-law tail is also pulsed, though to a smaller extent than the low-energy blackbody components. 

We performed a phase-resolved spectral analysis over the 0.3--10\,keV energy range using the EPIC-pn data. We extracted spectra from 50 phase intervals of width 0.02 rotational cycles, and fitted them using an absorbed double-blackbody model. In the fits, the column density was held fixed at the phase-averaged value ($\nh\ = 1.07\times10^{22}$\,\cmdue; Section\,\ref{sec:xrayanalysis}), while all other parameters were allowed to vary.
Figure\,\ref{fig:timing} shows the evolution of the spectral parameters and unabsorbed fluxes of both thermal components, as well as their flux ratio, along the rotational phase. The complex spin modulation pattern for the soft X-ray emission can be ascribed to changes in the blackbody radius of both components.
The smaller, hotter region traces more closely the fine structures seen in the pulse profile (e.g., the dip close to the peak; Figure\,\ref{fig:timing}). On the other hand, the scarce counting statistics available in the energy range 12--15\,keV (less than 500 net counts summing up the \nustar\ FPMs) precludes an assessment of possible variability of the power-law slope along the rotational phase.

\begin{figure*}
    \centering
    \includegraphics[width=2.12\columnwidth]{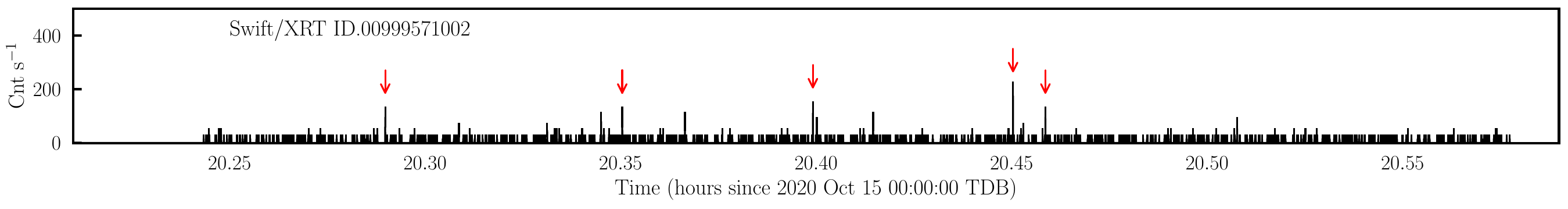}
    \includegraphics[width=2.12\columnwidth]{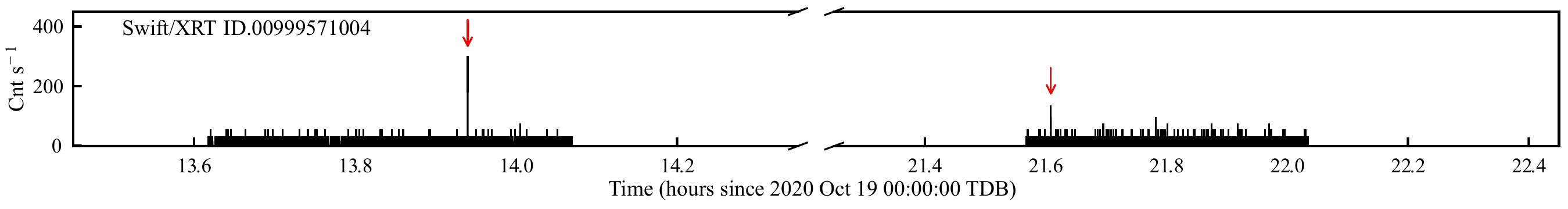}
    \includegraphics[width=2.12\columnwidth]{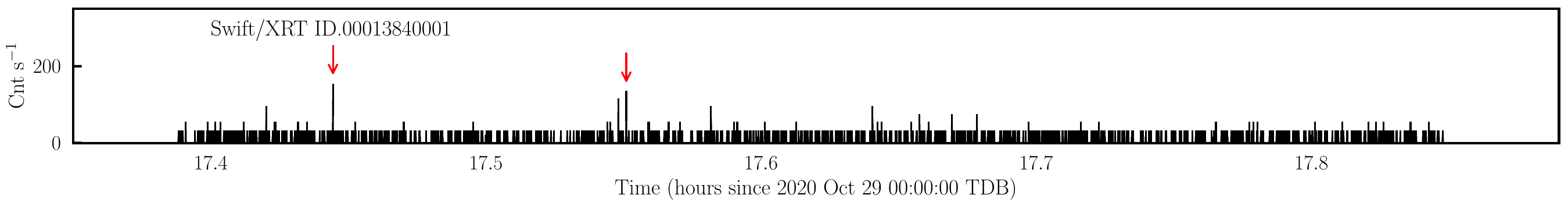}
    \includegraphics[width=2.12\columnwidth]{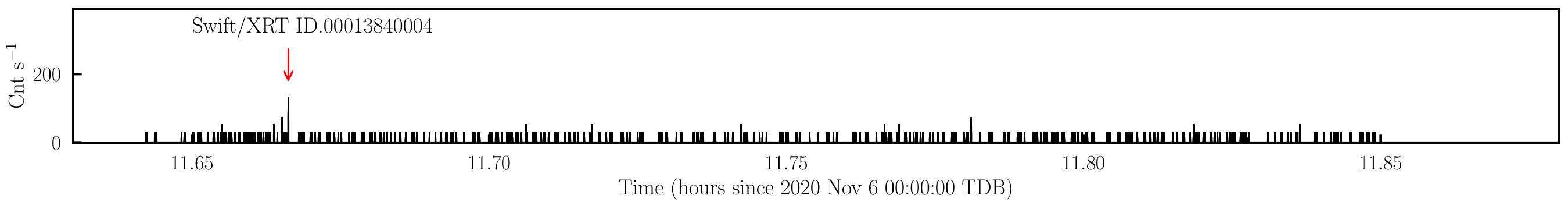}
    \includegraphics[width=2.12\columnwidth]{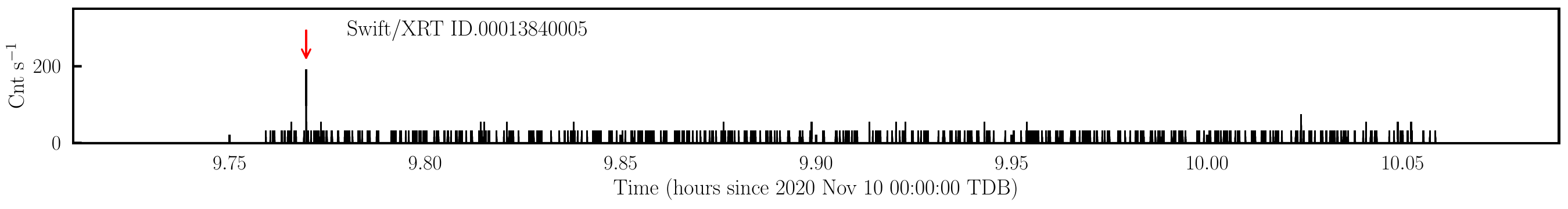}
    \includegraphics[width=2.12\columnwidth]{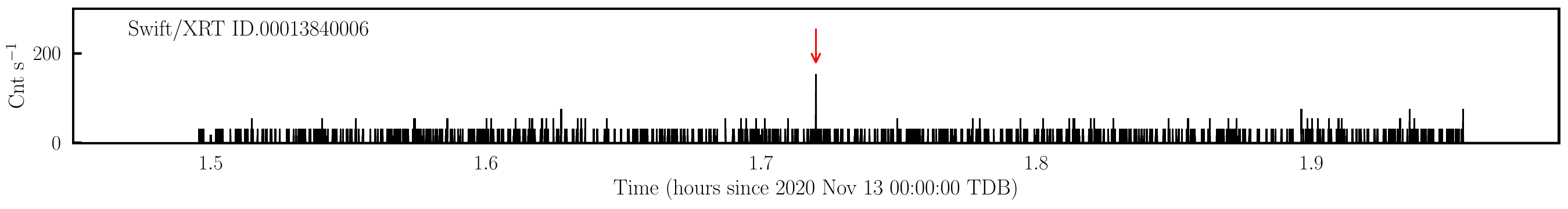}
    \caption{Light curves of \src\ extracted from the \swift/XRT data in which we detected bursts (0.3--10\,keV; time bin: 62.5\,ms). All events fulfilling our detection criterion are marked by arrows.}
    \label{fig:bursts}
\end{figure*}

\subsubsection{Search for Short X-Ray Bursts}
\label{sec:XRTbursts}
Short X-ray bursts were searched for by applying the procedure described by \citet{gkw04} and \citeauthor{scholz11} (\citeyear{scholz11}; see also \citealt{borghese20}). We extracted light curves with different time resolutions (2$^{-4}$, 2$^{-5}$, 2$^{-6}$\,s) to improve sensitivity to bursts of different durations, except for \swift/XRT PC-mode light curves that were binned at the timing resolution (2.5073\,s). For each observation, we calculated the Poisson probability of an event to be a random fluctuation with respect to the average number of counts per bin. Any bin with a probability smaller than $10^{-4}(N N_{\rm trials})^{-1}$, where $N$ is the total number of time bins and $N_{\rm trials}$ is the number of timing resolutions used in the search, was flagged as part of a burst. We detected 12 bursts in the \swift/XRT light curves (Figure\,\ref{fig:bursts} and Table\,\ref{tab:bursts}). No bursts were found in the \xmm\ and \nustar\ data sets.

\begin{deluxetable}{rrcccc}[htb]
\tablecaption{Log of X-Ray Bursts Detected in the \swift/XRT Data Sets
\label{tab:bursts}}
\tablecolumns{4}
\tablenum{2}
\tablewidth{0pt}
\tablehead{
\colhead{Obs.ID\tablenotemark{\scriptsize a}} &
\colhead{Burst Epoch} &
\colhead{Duration\tablenotemark{\scriptsize b}} &
\colhead{Fluence\tablenotemark{\scriptsize c}}\\ [-0.3cm]
\colhead{} & 
\colhead{YYYY Mmm DD hh:mm:ss (TDB)} &
\colhead{(ms)} & 
\colhead{(net counts)}
}
\startdata
00999571002 \#1 & 2020 Oct 15 20:17:23 & 62.5 & 6 \\
  \#2           &             20:21:01 & 62.5 & 6 \\    
  \#3           &             20:23:57 & 62.5 & 7 \\    
  \#4           &             20:27:01 & 62.5 & 11 \\                 
  \#5           &             20:27:31 & 62.5 & 6 \\ 
00999571004 \#1 & 2020 Oct 19 13:56:24 & 62.5 & 15 \\
  \#2           &             21:36:31 & 62.5 & 6 \\
00013840001 \#1 & 2020 Oct 29 17:26:40 & 62.5 & 7 \\
\#2             &             17:33:03 & 62.5 & 6 \\
00013840004 \#1 & 2020 Nov 6  11:39:58 & 62.5 & 6 \\
00013840005 \#1 & 2020 Nov 10 09:46:10 & 62.5 & 9 \\
00013840006 \#1 & 2020 Nov 13 01:43:12 & 62.5 & 7 \\
\enddata
\tablecomments{
\vspace{-0.25cm}
\tablenotetext{a}{The notation \#N corresponds to the burst number in a given observation.} \vspace{-0.25cm}
\tablenotetext{b}{The duration shall be considered as an approximate value. It was estimated as the coarser time resolution at which the burst is detected, or as the sum of the time bins showing enhanced emission for the case of the burst 00999571004 \#1.} \vspace{-0.25cm}
\tablenotetext{c}{The fluence refers to the 0.3--10\,keV energy range.} \vspace{-0.8cm}
}
\end{deluxetable}

\section{Radio Searches}\label{sec:radio}
Table\,\ref{tab:observations} reports a log of the radio observations, performed using the Sardinia Radio Telescope (SRT; \citealt{bolli,prandoni}) and \pks.

\subsection{Sardinia Radio Telescope Observations}\label{sec:srt} 
The SRT observed \src\ at 1.5\,GHz ($L$ band) on October 11 for 2.7\,hr and at 6.8\,GHz ($C$ band) on October 21, October 30 and November 6, for a total exposure of 22.5\,hr. 
Data were recorded with the ATNF PDFB backend in search mode over a bandwidth of 500\,MHz in $L$ band and 900\,MHz in $C$ band, with a spectral resolution of 1\,MHz. Total intensity data were 2 bit sampled every 0.1\,ms for the $L$ band and 0.125\,ms for the $C$ band.

All data were folded using the X-ray ephemeris and searched over a dispersion measure (DM) range of 0--1200\,pc\,cm$^{-3}$ and a spin period range of $\pm$1\,ms around the nominal value (after removing the most prominent radio frequency interference; RFI), using the software packages {\sc dspsr} \citep{vanstraten11} and {\sc psrchive} \citep{hotan04}. No pulsations were detected down to a folded signal-to-noise ratio of S/N=10. 
The corresponding flux density upper limits are reported in Table~\ref{tab:observations}. 

A search for bursts was performed on all data using the {\sc spandak} pipeline \citep{gajjar18}, sampling a DM range of 0--1200\,pc\,cm$^{-3}$. 
After dedispersion, the time series were searched for pulses using matched-filtering with a maximum window size of 32\,ms. After a first automatic sifting of the generated candidates, visual inspection was performed on the events that passed the selection. No bursts were found at either $L$ or $C$ band.

\subsection{\pks\ Observations}\label{sec:parkes}
The \pks\ radio telescope observed \src\ on October 12, starting at 08:04:24 UT for 2.9\,hr, simultaneously with \nustar.
Data were recorded with the ultra-wide-bandwidth low-frequency receiver (UWL; \citealt{hobbs20}) over a bandwidth of 3328\,MHz centered at 2368\,MHz. Full Stokes data were 4 bit sampled every 0.128\,ms. 
Four separate data sets covering different subbands were created with different spectral resolutions so as to achieve a maximum broadening of a few ms for a signal with DM=600\,pc\,cm$^{-3}$ \footnote{This is the maximum value expected for our Galaxy in the direction of \src, according to the NE2001 electron density model \citep{cordes02}.} (band b0, from 704 to 1216\,MHz, split into 2048 frequency channels; b1, 1216--1984\,MHz, 768 channels; b2, 1984--2752\,MHz, 384 channels; b3, 2752--4032\,MHz, 320 channels).

To search for persistent pulsations, the data of the subbands were folded using the X-ray ephemeris. After RFI removal, the frequency resolution of the folded data was scaled uniformly down to 4\,MHz, and data were summed together. A search over a DM range of 0--1200\,pc\,cm$^{-3}$ spanning $\pm$1\,ms around the nominal spin period was performed on the entire observing bandwidth. No pulsations were detected down to a folded S/N=10.
The flux density upper limits in the different subbands are reported in Table\,\ref{tab:observations}.

The single-pulse analysis was performed using the procedure described in Section\,\ref{sec:srt} and the same parameters. The subbanding did not affect the search since we are looking for impulsive bursts. After the RFI removal, the data were dedispersed from 0 to 1200\,pc\,cm$^{-3}$. No bursts were found at any band.

\begin{figure*}
  \centering
    \includegraphics[width=2.25\columnwidth]{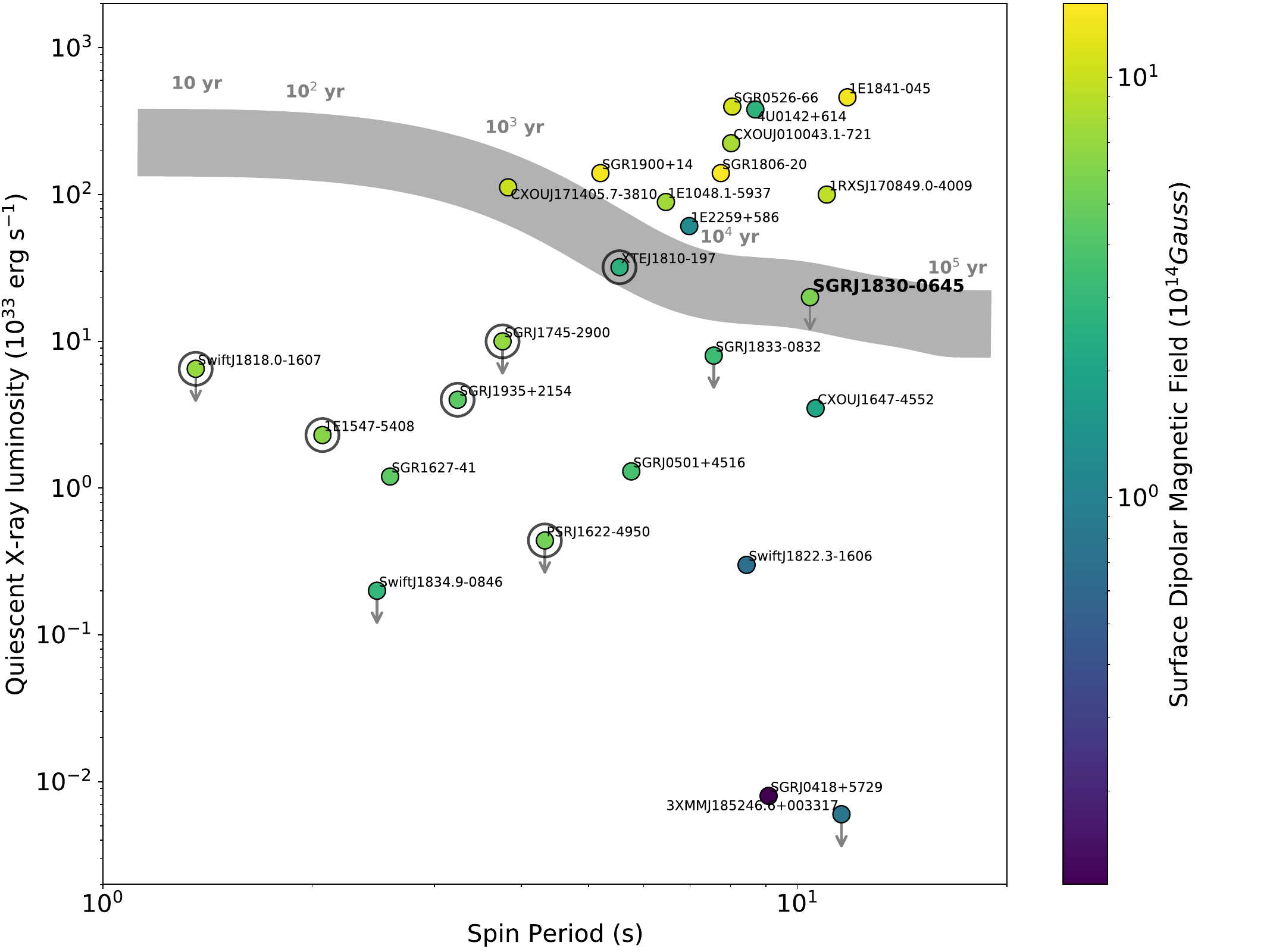} 
    \caption{Quiescent X-ray luminosity as a function of the spin period for magnetars, including \src\ (in bold). Circles mark radio-loud magnetars. The gray shaded region shows the magneto-thermal evolutionary path of \src\ according to the model discussed in the text. Values are taken from\dataset[The Magnetar Outburst Online Catalog]{http://magnetars.ice.csic.es/}.}
    \label{fig:discussion}
\end{figure*}

\section{Discussion}\label{sec:discussion}
In 2020 October, \src\ entered its first detected outburst phase, revealing to be a new Galactic magnetar. The source emitted numerous X-ray bursts since the beginning of the outburst and during our monitoring campaign (Sections\,\ref{sec:BATbursts} and \ref{sec:XRTbursts}; see also \citealt{palmer20,ray20}). This is a distinctive phenomenology usually observed in magnetars during an active phase.
Our campaign allowed us to measure the spin period ($P = 10.415724(1)$\,s) and its first derivative ($\dot{P} = 7(1)\times10^{-12}$\,s\,s$^{-1}$) at the outburst peak (Section\,\ref{sec:timing}), hence to estimate a surface dipolar magnetic field $B_{\rm{dip}} \sim 6.4\times10^{19}(P\dot{P})^{1/2} \approx5.5\times10^{14}$\,G at pole (using the vacuum dipole formula), a spin-down luminosity $\dot{E}_{\rm {rot}} = 4\pi^2 I\dot{P} P^{-3} \approx2.4\times10^{32}$\,\lum\ (where $I\approx10^{45}$\,g\,cm$^2$ is the moment of inertia of the NS) and a characteristic age $\tau_c=P/2\dot{P}\approx24$\,kyr. 

The extension of the Galaxy in the direction of \src\ estimated from the maps of \citet{hou14} and the similarity between the column density derived from our spectral fits ($\nh\sim1.07\times10^{22}$\,cm$^{-2}$) and that expected in the direction of the source within the Galaxy ($N_{\rm H,Gal}\sim1.1\times10^{22}$\,cm$^{-2}$; \citealt{willingale13}) suggest a distance $D$$\gtrsim$5\,kpc. In the following, we rescale all quantities to a distance of $D$$=$10\,kpc.
Assuming isotropic emission, the X-ray luminosity at the outburst peak is then $L_{{\rm X,p}}\sim6\times10^{35} d^2_{10}$\,\lum, while the limit that we derived for the quiescent X-ray luminosity is $L_{{\rm X,q}}<2\times10^{34} d^2_{10}$\,\lum\ (0.3--10\,keV; $d_{10} = D/10\,{\rm kpc}$; see Section\,\ref{sec:xrayanalysis}).

We studied the evolutionary history of \src\, using a two-dimensional magneto-thermal evolutionary model \citep{vpm12,vigano13,vigano20}. We used crustal-confined models consisting of an initial poloidal dipolar field ($B_{\rm dip,in}$) plus a toroidal field ($B_{\rm tor,in}$), and assumed for simplicity an equal amount of magnetic energy in the two components. We find that the evolution of an initial configuration with $B_{\rm dip,in}\sim10^{15}$\,G and $B_{\rm tor,in}\sim10^{16}$\,G provides a close match to the current $P$ and $\dot{P}$ after $\sim$23\,kyr. We thus obtain an age $\tau_{\rm th}\sim23$\,kyr for \src, similar to its characteristic age, and predict a quiescent bolometric thermal luminosity at $\tau_{\rm th}$ that would be just below our current upper limit on $L_{{\rm X,q}}$. The good agreement between $\tau_{\rm th}$ and $\tau_{\rm c}$ is due to the fact that the dissipation of the magnetic field is not yet substantial at this evolutionary stage. At later stages, $\tau_{\rm c}$ will overestimate the real age because the electromagnetic torque decreases in time due to magnetic field dissipation, while $\tau_{\rm th}$ usually remains consistent with the real age.  
 
From its properties and simulated history, we find that \src\ is a middle age magnetar that had a relatively strong magnetic energy at birth. The gray shaded region in Figure\,\ref{fig:discussion} shows the evolution of the spin period and luminosity of \src\ (taking into account uncertainties due to the assumption of light or heavy elements in the envelope), compared with current values for other magnetars. This region encompasses also two other magnetars with rather different magnetic field strengths, \object[CXOU J171405.7-381031]{\ctb} and \object[XTE J1810-197]{\xte} (see the color scale in the figure). The properties of \src\ at birth were probably similar to those of \ctb, but different from those of \xte. Indeed, \ctb\ has a stronger magnetic field and a younger age than \src, and it might be now at an earlier stage of a similar evolutionary scenario. On the other hand, \xte\ is characterized by a luminosity and spin period that might be potentially compatible with those expected if this source were at an earlier stage of the same evolutionary path of \src, but its magnetic field is already smaller than that of \src.

The X-ray emission properties observed from \src\ soon after the outburst onset are in line with those of other magnetars (\citealt{cotizelati18} and references therein), and fit well within the resonant Compton scattering scenario (RCS; \citealt{tlk02,rea08,wadiasingh18}; see also \citealt{turolla15} and references therein). The dominance of the thermal over the nonthermal (power-law) component (Section\,\ref{sec:xrayanalysis}) suggests that the magnetospheric twist is restricted to a bundle of current-carrying field lines. Ohmic dissipation of the returning currents on the star surface leads to the appearance of localized hot spots \citep{beloborodov09}. Heated regions may form on the surface also as a consequence of heat released deeper in the crust by local dissipation of magnetic energy \citep{pons12}. In this respect, our analysis suggests the existence of two heated regions of different temperatures and sizes on the surface of \src: an extended warm region (average blackbody temperature $\sim$0.45\,keV and radius $\sim$6\,km), and a small hot region (temperature $\sim$1.1\,keV and radius $\sim$1.5\,km). According to our analysis, the modulation of the light curve is (almost) entirely due to the change of the visible area of these regions, the temperatures being fairly constant along the spin phase (Figure\,\ref{fig:timing}). The phase-alignment between the light-curve profiles associated with the two blackbody components is indicative of a scenario in which the hot and the warm regions are not spatially separated. A more plausible picture is that what we actually see is a single heated spot with a complex shape where an extended warm region is surrounded by a smaller hotter region with an asymmetric structure (the former producing the relatively broad peak observed in the light-curve profile, the latter resulting in the two prominent narrow peaks separated by $\sim$0.2 rotational phase cycles; see Figure\,\ref{fig:timing}). Indeed, recent 3D simulations have shown that heat injected in a localized region of the crust of a magnetar flows anisotropically to the surface, leading to the appearance of a hot spot with a complex shape and a nonuniform temperature distribution \citep{degrandis20}. Soft, thermal photons coming from such a spot can produce a complex pulse profile with a high pulsed fraction, which would be smaller ($\lesssim$20\%) in the case of two circular, antipodal hot spots \citep{albano10,turolla13a}.
The pulsed fraction decrease in going from lower to higher energies (up to $\sim$25\,keV) is different from what is observed in other magnetars \citep[see, e.g.,][and references therein]{an15}. Different trends of the pulsed fraction variability with energy can be envisaged within the RCS scenario, depending on the viewing angles as well as the location and velocity distribution of the charged particles in the magnetosphere that up-scatter thermal photons. In the case of \src, the RCS mechanism tends to wash out the imprint of the (pulsed) primary emission.

Our nondetection of radio pulsations from \src\ is not that surprising per se. Similarly to canonical radio pulsars, radio-loud magnetars are generally characterized by high spin-down luminosities $\dot{E}_{\rm {rot}} \gtrsim 10^{33}$\,\lum, implying $L_{{\rm X,q}}/\dot{E}_{\rm {rot}}<1$ (\citealt{rea12}; the only outlier being
\object[XTE J1810-197]{\xte}). A reliable estimate of $L_{{\rm X,q}}/\dot{E}_{\rm {rot}}$ for \src\ is difficult owing to uncertainties on the distance and the nondetection of the source in pre-outburst X-ray data. 
However, if its quiescent luminosity is not much below our quoted limit, it would have $L_{{\rm X,q}}/\dot{E}_{\rm {rot}}\gg1$, in line with being radio-silent. 
Alternatively, \src\ might be radio-loud but undetectable due to unfavorable beaming.

Future high-sensitive X-ray observations will be key in mapping the evolution of the heated spot on the star surface and the long-term contribution of magnetospheric currents to the broadband X-ray emission of \src. \\

We thank N.~Schartel and F.~Harrison for approving Target of Opportunity observations with \xmm\ and \nustar\ in the Director's Discretionary Time, and the \xmm\ and \nustar\ SOCs for carrying out the observations. We also thank B.~Cenko and the \swift\ duty scientists and science planners for making the \swift\ Target of Opportunity observations possible.
This research is based on observations with \xmm\ (ESA/NASA), \nustar\ (CalTech/NASA/JPL), \swift\ (NASA/ASI/UKSA) and on data retrieved through the NASA/GSFC's HEASARC archives. 
The Sardinia Radio Telescope is funded by the Italian MIUR, ASI, and the Autonomous Region of Sardinia, and is operated as a National Facility by INAF. We used data collected at the Parkes radio telescope (proposal No. P1083), part of the Australia Telescope National Facility which is funded by the Australian Government for operation as a National Facility managed by CSIRO. We acknowledge the Wiradjuri people as the traditional owners of the Observatory site.
F.C.Z. and A.B. are supported by Juan de la Cierva fellowships. F.C.Z., A.B., N.R., C.D., and D.V. are supported by the ERC Consolidator Grant ``MAGNESIA" (No. 817661) and acknowledge funding from grants SGR2017-1383 and PGC2018-095512-BI00. G.L.I., P.E., R.T., and A.T. acknowledge financial support from the Italian MIUR through PRIN grant 2017LJ39LM. G.L.I. also acknowledges funding from ASI-INAF agreements I/037/12/0 and 2017-14-H.O.
M.B., A.P., and A.R. acknowledge financial support from the research grant ``iPeska'' (PI: A.~Possenti) funded under the INAF national call PRIN-SKA/CTA with Presidential Decree 70/2016. A.R. acknowledges continuing valuable support from the Max-Planck Society. We acknowledge the support of the PHAROS COST Action (CA16214). \vspace{-0.2cm} 
\facilities{\xmm\ (EPIC), \swift\ (BAT, XRT), \nustar, \rst\ (PSPC), \pks, SRT, ADS, HEASARC.} \vspace{-0.6cm}
\software{HEASOFT (v.6.28; \citealt{heasoft14}), NUSKYBGD \citep{wik14}, SAOImageDS9 (v.8.1; \citealt{joye03}), SAS (v.19; \citealt{gabriel04}), XSPEC (v.12.11.1; \citealt{arnaud96}), DSPSR \citep{vanstraten11}, PSRCHIVE \citep{hotan04}, SPANDAK (\url{https://github.com/gajjarv/PulsarSearch}), Astropy (v.4.2; \citealt{astropy:2013, astropy:2018}), MATPLOTLIB (v3.3.3; \citealt{hunter07}), NUMPY (v1.19.4; \citealt{harris20}).}
 
\section*{ORCID iDs}
\noindent 
F. Coti Zelati\hskip2pt\includegraphics[width=7pt]{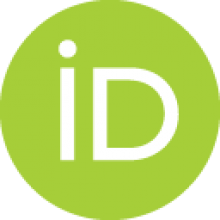} \url{https://orcid.org/0000-0001-7611-1581}\\
A. Borghese\hskip2pt\includegraphics[width=7pt]{orcid-ID.png} \url{https://orcid.org/0000-0001-8785-5922}\\
G.~L. Israel\hskip2pt\includegraphics[width=7pt]{orcid-ID.png} \url{https://orcid.org/0000-0001-5480-6438}\\
N. Rea\hskip2pt\includegraphics[width=7pt]{orcid-ID.png} \url{https://orcid.org/0000-0003-2177-6388}\\  
P. Esposito\hskip2pt\includegraphics[width=7pt]{orcid-ID.png} \url{https://orcid.org/0000-0003-4849-5092}\\  
M. Pilia\hskip2pt\includegraphics[width=7pt]{orcid-ID.png} \url{https://orcid.org/0000-0001-7397-8091}\\  
M. Burgay\hskip2pt\includegraphics[width=7pt]{orcid-ID.png} \url{https://orcid.org/0000-0002-8265-4344}\\  
A. Possenti\hskip2pt\includegraphics[width=7pt]{orcid-ID.png} \url{https://orcid.org/0000-0001-5902-3731}\\  
A. Corongiu\hskip2pt\includegraphics[width=7pt]{orcid-ID.png} \url{https://orcid.org/0000-0002-5924-3141}\\  
A. Ridolfi\hskip2pt\includegraphics[width=7pt]{orcid-ID.png} \url{https://orcid.org/0000-0001-6762-2638}\\  
C. Dehman\hskip2pt\includegraphics[width=7pt]{orcid-ID.png} \url{https://orcid.org/0000-0003-0554-7286}\\  
D. Vigan\`o\hskip2pt\includegraphics[width=7pt]{orcid-ID.png} \url{https://orcid.org/0000-0001-7795-6850}\\  
R. Turolla\hskip2pt\includegraphics[width=7pt]{orcid-ID.png} \url{https://orcid.org/0000-0003-3977-8760}\\  
S. Zane\hskip2pt\includegraphics[width=7pt]{orcid-ID.png} \url{https://orcid.org/0000-0001-5326-880X}\\  
A. Tiengo\hskip2pt\includegraphics[width=7pt]{orcid-ID.png} \url{https://orcid.org/0000-0002-6038-1090}\\  
E.~F. Keane\hskip2pt\includegraphics[width=7pt]{orcid-ID.png} \url{https://orcid.org/0000-0002-4553-655X}\\  

\bibliographystyle{aasjournal}

{}

\end{document}